\documentclass[aps,prl,reprint,twocolumn,superscriptaddress,nofootinbib]{revtex4-1}

\usepackage{amsthm}
\usepackage{amsmath}
\usepackage{graphicx}
\usepackage{slashed}
\usepackage{amssymb}
\usepackage{float}
\usepackage[colorlinks=True, citecolor=blue, urlcolor=blue, linkcolor=blue]{hyperref}

\newcommand{\nn}{\nonumber}

\newcommand{\be}{\begin{eqnarray}}
\newcommand{\ee}{\end{eqnarray}}

\newcommand{\gann}{\textsc{GAN}}
\newcommand{\ganns}{\textsc{GAN}s}
\newcommand{\dglap}{\textsc{DGLAP}}

\begin{document}

\title{Explainable machine learning of the underlying physics of high-energy \\ particle collisions}

\author{Yue Shi Lai}
\email{ylai@lbl.gov}
\affiliation{Nuclear Science Division, Lawrence Berkeley National Laboratory, Berkeley, California 94720, USA}

\author{Duff Neill}
\email{duff.neill@gmail.com}
\affiliation{Theoretical Division, MS B283, Los Alamos National Laboratory, Los Alamos, NM 87545, USA}

\author{Mateusz P\l osko\'n}
\email{mploskon@lbl.gov}
\affiliation{Nuclear Science Division, Lawrence Berkeley National Laboratory, Berkeley, California 94720, USA}

\author{Felix Ringer}
\email{fmringer@lbl.gov}
\affiliation{Nuclear Science Division, Lawrence Berkeley National Laboratory, Berkeley, California 94720, USA}

\date{\today}

\begin{abstract}
We present an implementation of an explainable and physics-aware machine learning model capable of inferring the underlying physics of high-energy particle collisions using the information encoded in the energy-momentum four-vectors of the final state particles. We demonstrate the proof-of-concept of our White Box AI approach using a Generative Adversarial Network (GAN) which learns from a DGLAP-based parton shower Monte Carlo event generator. We show, for the first time, that our approach leads to a network that is able to learn not only the final distribution of particles, but also the underlying parton branching mechanism, i.e. the Altarelli-Parisi splitting function, the ordering variable of the shower, and the scaling behavior. While the current work is focused on perturbative physics of the parton shower, we foresee a broad range of applications of our framework to areas that are currently difficult to address from first principles in QCD. Examples include nonperturbative and collective effects, factorization breaking and the modification of the parton shower in heavy-ion, and electron-nucleus collisions.
\end{abstract}

\maketitle

{\it Introduction.} In recent years machine learning techniques have lead to range of new developments in nuclear and high-energy physics~\cite{deOliveira:2015xxd,Komiske:2016rsd,Kasieczka:2017nvn,Metodiev:2017vrx,Englert:2018cfo,Hashemi:2019fkn,Otten:2019hhl,Butter:2019cae,DiSipio:2019imz,Farrell:2019fsm,Alanazi:2020klf,Pang:2016vdc,Komiske:2017ubm,Ball:2017nwa,Paganini:2017hrr,Datta:2017lxt,Larkoski:2017jix,Chien:2018dfn,Collins:2018epr,Zhou:2018ill,Lai:2018ixk,Komiske:2018cqr,Du:2019civ,Pang:2019aqb,Andreassen:2019cjw,Carrazza:2019efs,Kasieczka:2020nyd,Li:2020vav,Kanwar:2020xzo}. For example, in Refs.~\cite{deOliveira:2015xxd,Komiske:2016rsd,Kasieczka:2017nvn,Metodiev:2017vrx,Englert:2018cfo} jet tagging techniques were developed which often outperform traditional techniques. In Refs.~\cite{Hashemi:2019fkn,Otten:2019hhl,Butter:2019cae,DiSipio:2019imz,Farrell:2019fsm,Alanazi:2020klf} Generative Adversarial Networks (\ganns)~\cite{Goodfellow:2014,Radford2016UnsupervisedRL}, a form of unsupervised machine learning, were used to simulate event distributions in high-energy particle collisions. There have also been efforts to infer physics information from data. In Ref.~\cite{Andreassen:2018apy} a probabilistic model was introduced based on jet clustering and in Ref.~\cite{Monk:2018zsb} a convolutional autoencoder within a shower was used which qualitatively reproduces jet observables. See also Refs.~\cite{Bogatskiy:2020tje,Larkoski:2020thc,Faucett:2020vbu} for recent work on physics-aware learning.

The underlying physics information of high-energy particle collisions is encoded in hard-scattering processes, the subsequent parton shower and the hadronization mechanism. These steps are modeled by general purpose parton showers used in Monte Carlo event generators which play an important role in our understanding of high-energy collider experiments~\cite{Sjostrand:2007gs,Bahr:2008pv,Gleisberg:2008ta}. Starting with highly energetic quarks or gluons which are produced in hard-scattering events, parton showers simulate the parton branching processes that occur during the evolution from the hard scale to the infrared which is followed by the hadronization step. While the general concept of parton showers is well established, important questions about the perturbative accuracy~\cite{Nagy:2012bt,Hoche:2015sya,Alioli:2015toa,Dasgupta:2018nvj,Bewick:2019rbu,Dasgupta:2020fwr,Forshaw:2020wrq}, nonperturbative effects~\cite{Andersson:1983ia,Marchesini:1987cf,Metz:2016swz,Neill:2020mtc} and the modification in the nuclear environment~\cite{Gyulassy:1993hr,Baier:1996sk,Zakharov:1996fv,Gyulassy:2000er,Wang:2001ifa,Arnold:2002ja,Qiu:2004da,Liu:2006ug,Armesto:2011ht,Mehtar-Tani:2013pia,Burke:2013yra,Qiu:2019sfj,Putschke:2019yrg,Caucal:2019uvr,Vaidya:2020cyi}, remain a challenge.

\begin{figure}[t]
\vspace{-4cm}
\hspace*{-.5cm}
\includegraphics[scale=0.4]{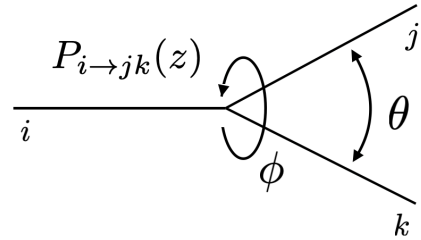}
\vspace{-4cm}
\caption{Parton splitting process $i\to jk$ with longitudinal momentum fraction $z$, relative splitting angle of the two daughter partons $\theta$ and azimuthal angle $\phi$.~\label{fig:splitting}}
\end{figure}
In this work, we propose an explainable or White Box AI approach~\cite{NEURIPS2018_842424a1,DBLP:journals/corr/abs-1811-12530} to learn the underlying physics of high-energy particle collisions. As a proof of concept, we present results of a \gann\ trained on the final output of a parton shower, which not only reproduces the final distribution of particles but also learns the underlying showering mechanism using the complete event information. 
\begin{figure*}[t]
  \centerline{\hfill \includegraphics[width=\textwidth]{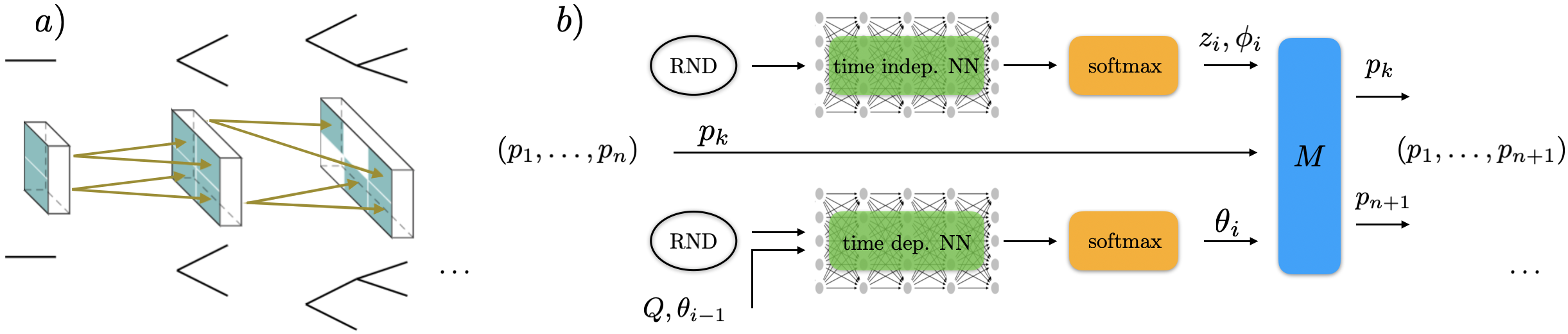}\hfill}
  \caption{Schematic illustrations of the generator network: $a)$ Parallelized data structure of the random splitting trees executed on the GPU. $b)$ Flow diagram of the $i$th splitting process ($n\to n+1$ partons) of a randomly chosen parton with momentum $p_k$. The time dependent and independent networks are shown which take as input random numbers (RND) as well as $Q,\theta_{i-1}$ in the time dependent case. The output of the two neural networks is passed through a softmax to the module $M$ which determines the four-vectors of the two daughter partons from the variables of the $1\to 2$ splitting process and the parent momentum $p_k$.~\label{fig:generator}}
\end{figure*}
\ganns\ consist of two competing neural networks, the generator and discriminator. The design of our generator network allows to not only describe the final distribution of particles of the shower but the different layers also give access to the underlying physics encoded in the parton branching processes. More specifically, we demonstrate that the network can learn the Altarelli-Parisi splitting function $P_{i\to jk}(z)$, the splitting angles of individual branching processes and the dependence of the shower on the energy scale $Q$, see Fig.~\ref{fig:splitting}. This is achieved by separating the \gann\ into two components such that it can learn both self-similar/fractal aspects of the shower like the Altarelli-Parisi splitting function as well as Monte Carlo time dependent variables such as the splitting angle. We use a network architecture that is sufficiently general, and as a result, capable of incorporating nonperturbative physics in the future. In order to use the complete information of each event, we use data representation which is directly given by the four-vectors of the final state particles. To avoid sensitivity to the unphysical ordering of the list of four-vectors during the training process, we use sets to represent the data. In particular, in our work, the necessary permutation invariance is achieved by using so-called deep sets which were developed in Refs.~\cite{DBLP:journals/corr/ZaheerKRPSS17,DBLP:journals/corr/abs-1901-09006,JMLR:v21:19-322}. 

With the framework introduced in this work, we can access the underlying physics mechanisms effectively departing from the typical black-box paradigm for neural networks. Moreover, we expect that eventually the \gann\ can be trained directly on experimental data (i.e. measured four-vectors of detected particles). Generally, \ganns\ are ideally suited for such applications due to their generalizability and robustness when exposed to imperfect data sets. We expect that our approach will be particularly relevant for studies of heavy-ion collisions at RHIC and the LHC as well as electron-nucleus collisions at the future Electron-Ion Collider~\cite{Accardi:2012qut}. In heavy-ion collisions, the presence of quark-gluon plasma (QGP)~\cite{PhysRevD.27.140, Arsene:2004fa,Adcox:2004mh,Back:2004je,Adams:2005dq,Jacak:2012dx,LHC1review,Braun-Munzinger:2015hba, TheBigPicture} leads to modifications of highly energetic jets as compared to the proton-proton baseline. These phenomena are typically referred to as \emph{jet quenching}. Significant theoretical~\cite{Gyulassy:1993hr,Baier:1996sk,Zakharov:1996fv,Gyulassy:2000er,Wang:2001ifa,Arnold:2002ja,Qiu:2004da,Liu:2006ug,Armesto:2011ht,Burke:2013yra,Qiu:2019sfj,Putschke:2019yrg,Vaidya:2020cyi} and experimental~\cite{Adare:2010de,Sirunyan:2017isk,Adamczyk:2017yhe,Acharya:2019jyg,Aaboud:2018twu} efforts have been made to better understand the physics of this process. Using the novel techniques proposed in this work, we will eventually be able to analyze the properties of the medium modified parton shower using, for the first time, the complete event information.

{\it The parton shower.} The parton shower we use for training the \gann\ is designed to solve the \dglap\ evolution equations, see Refs.~\cite{Dasgupta:2014yra,Neill:2020mtc}. In addition, we set up the full event kinematics in spherical coordinates such that we can use the final distribution of partons generated by the shower as input to the adversarial training process. We start with a highly energetic parton which originates from a hard-scattering event at the scale $Q$. The parton shower cascade is obtained through recursive $1\to 2$ branching processes according to the \dglap\ evolution equations. There are three variables that describe a \dglap\ splitting process $i\to jk$ as illustrated in Fig.~\ref{fig:splitting}. First, the large light cone momentum fraction $z$ of the daughter partons relative to the parent is determined by sampling from the Altarelli-Parisi splitting functions. Second, the orientation of the two daughter partons, the azimuthal angle $\phi$, is obtained by sampling from a flat distribution in the range $[-\pi,\pi]$. Third, the splitting angle $\theta$ which is the relative opening angle of the two daughter partons, is determined as follows: First, sample a Monte Carlo time step $\Delta t$ from the no-emission Sudakov factor
\begin{equation}\label{eq:Deltat}
    \exp\Bigg[-\Delta t \sum_{i=q,\bar q, g}\int\limits_\epsilon^{1-\epsilon}{\rm d}z\, P_i(z)\Bigg]\,,
\end{equation}
where the $P_i$ denote the final state summed Altarelli-Parisi splitting functions for (anti-)quarks and gluons. Then advance the shower time $t\to t+\Delta t$ and solve for the splitting angle $\theta$ in
\begin{equation}\label{eq:MCtime}
    t(Q,\theta)= \int\limits_{Q\tan(\pi/2)}^{Q\tan(\theta/2)}\frac{{\rm d}t'}{t'}\frac{\alpha_s(t')}{\pi}\,.
\end{equation}
We evolve the shower from the hard scale $Q$ down to the hadronization scale which we choose as 1~GeV. We note that the \dglap\ shower described here has two cutoff parameters. First, the angular cutoff on the splitting angle $\theta$ which is introduced by the hadronization scale and which determines the end of the shower. Second, we introduce the cutoff $\epsilon$ on the momentum fraction $z$, see Eq.~(\ref{eq:Deltat}). For our numerical results we choose $\epsilon=0.03$ which avoids the singular endpoints. The generated spectrum is accurate in the range $\epsilon < z < 1-\epsilon$, and emitted partons that violate these bounds are not evolved further in the shower. 
\begin{figure*}[!t]
  \centerline{
  \includegraphics[width = 0.3333 \textwidth]{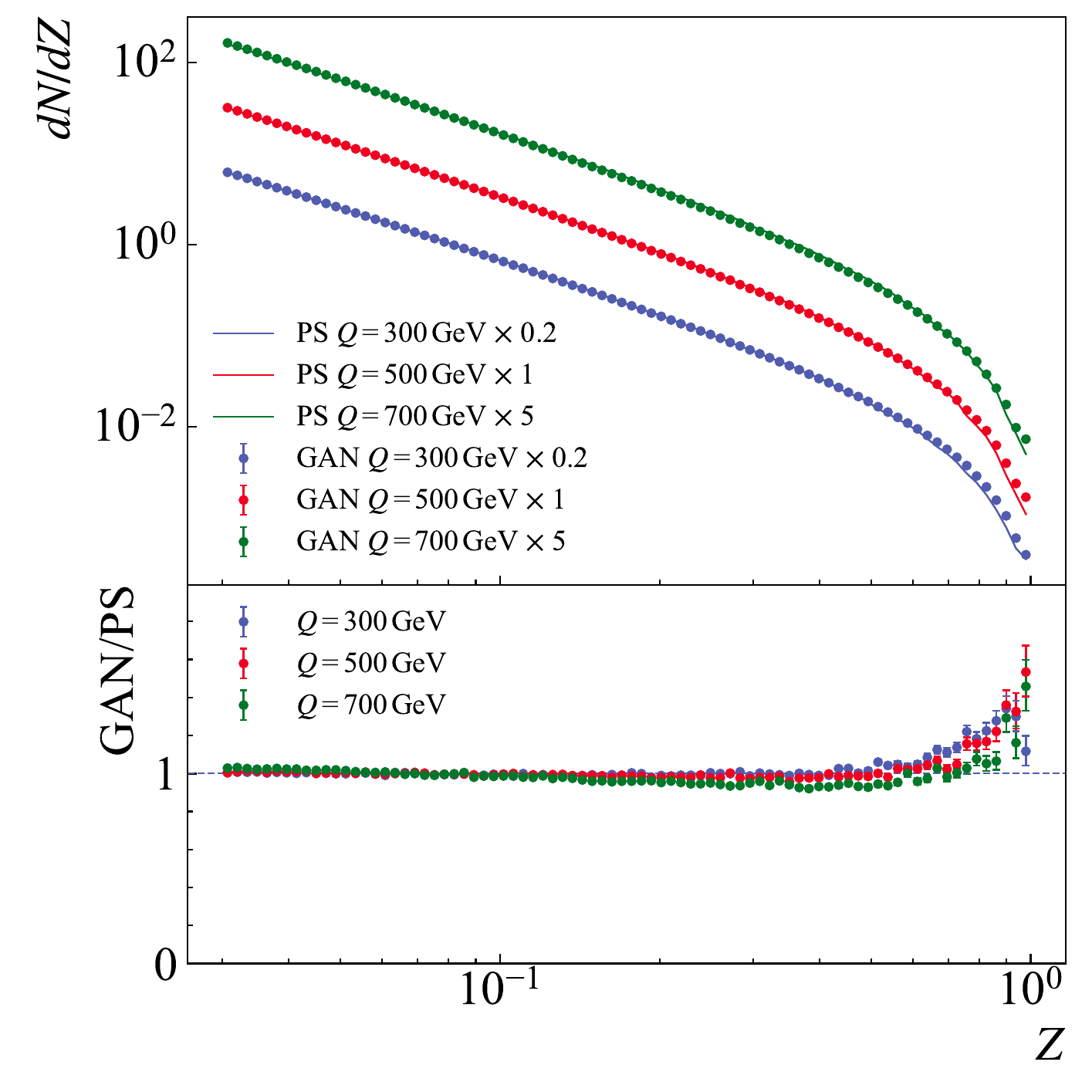}\hfill
  \includegraphics[width = 0.3333 \textwidth]{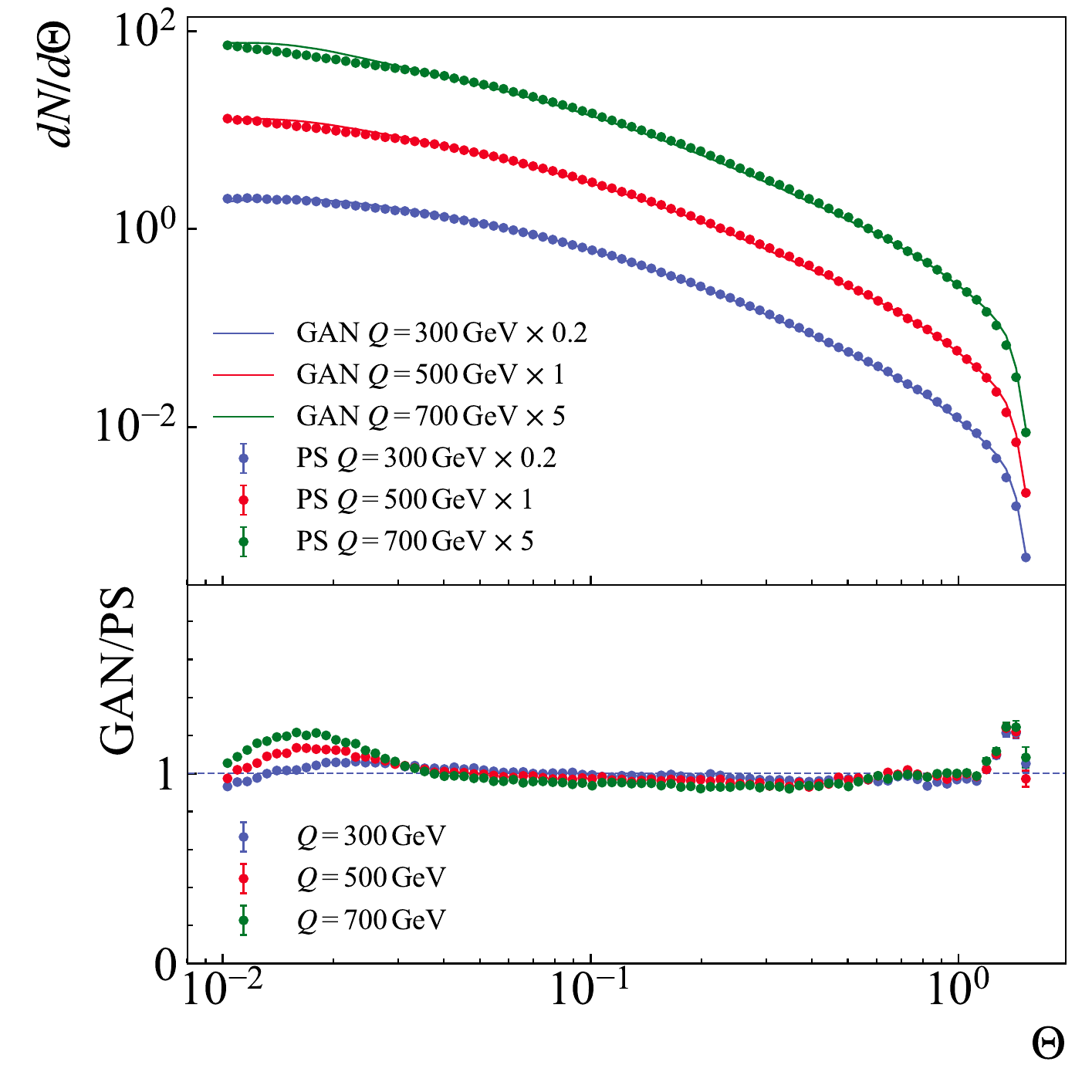}\hfill
  \includegraphics[width = 0.3333 \textwidth]{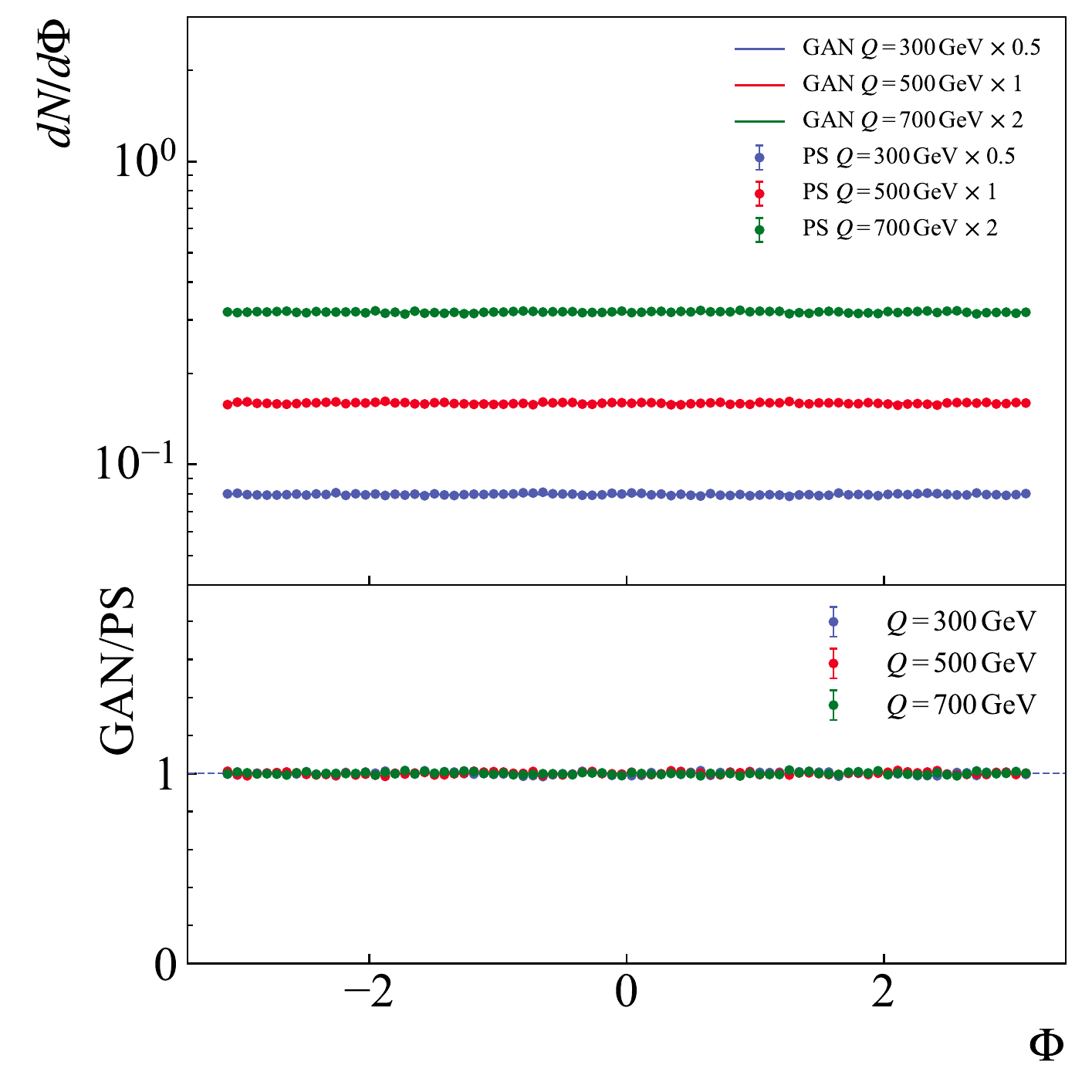}}
  \caption{Comparison of the parton shower and \gann\ in terms of the final distribution of particles. The three panels show the momentum fraction $Z$, the polar angle $\Theta$ and the azimuthal angle $\Phi$ (from left to right) for $Q=300,500,700$~GeV.~\label{fig:final_particle_dist}}
\end{figure*}
From the parent direction and the variables $(z,\theta,\phi)$ of a given $1\to 2$ splitting, we set up the full event kinematics and determine the absolute position of the two daughter partons in spherical coordinates $(\tilde \Theta,\tilde \Phi)$. The relevant kinematic relations are summarized in the supplemental material. After the shower terminates, we record the final momentum fractions $Z$ of the partons relative to the initial momentum scale $Q$ as well as their corresponding spherical coordinates $(\Theta,\Phi)$\footnote{Note that we use the variables $(z,\theta,\phi)$ to describe an individual $1\to 2$ splitting processes as shown in Fig.~\ref{fig:splitting}, $(\tilde \Theta,\tilde \Phi)$ are the spherical coordinates of partons at intermediate stages of the shower and $(Z,\Theta,\Phi)$ denote the final distributions of the momentum fraction and angles of the partons after the shower terminates.}. Together with the on-shell condition they fully specify the exclusive final state distribution of all particles which are produced by the shower. We note that the variables $z,\phi$ are independent of the shower time $t$ (self-similar or fractal variables), whereas the splitting angle $\theta$ is determined from the ordering variable of the shower and it also depends on the scale $Q$. Therefore, we treat $\theta$ differently from the other two variables in the generator network, as discussed below. The shower described here provides an ideal test ground to explore the use of explainable machine learning that aims to extract the structure of the parton shower, and thus the underlying physics, from the final distribution of particles in the event. We leave the investigation of other shower algorithms and nonperturbative effects for future work. 

\begin{figure*}[!t]
  \centerline{
  \includegraphics[width = 0.33 \textwidth]{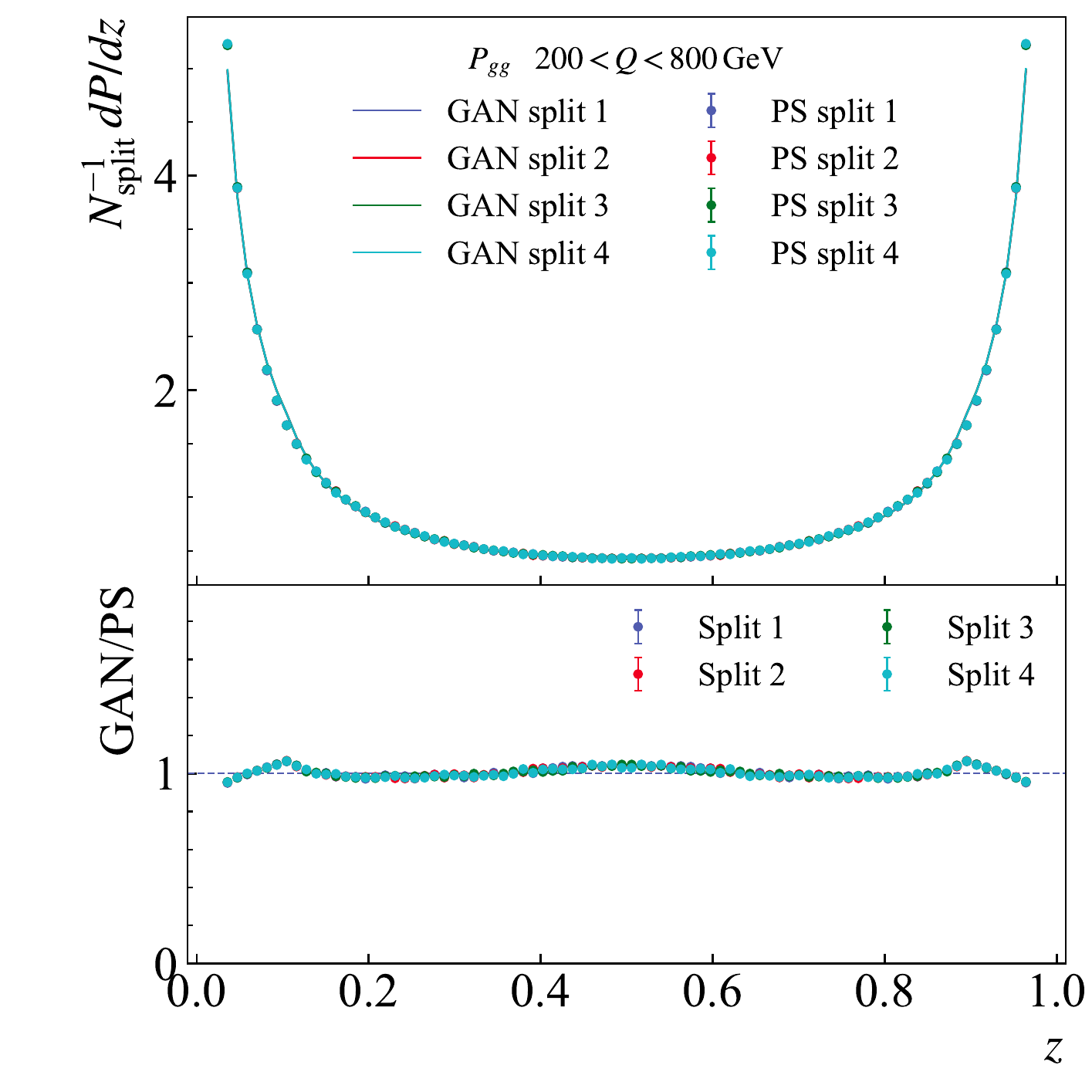}\hfill
  \includegraphics[width = 0.33 \textwidth]{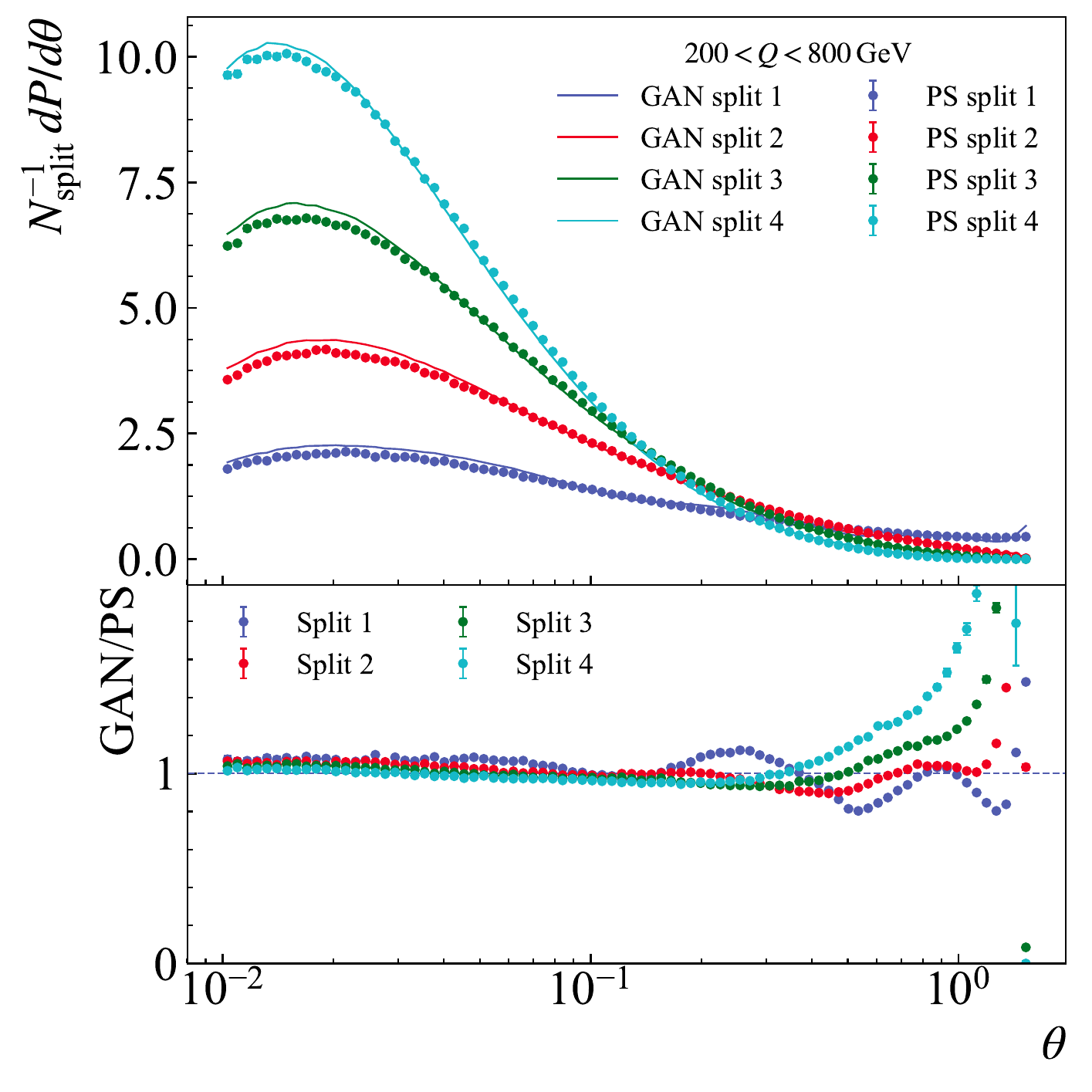}\hfill
  \includegraphics[width = 0.33 \textwidth]{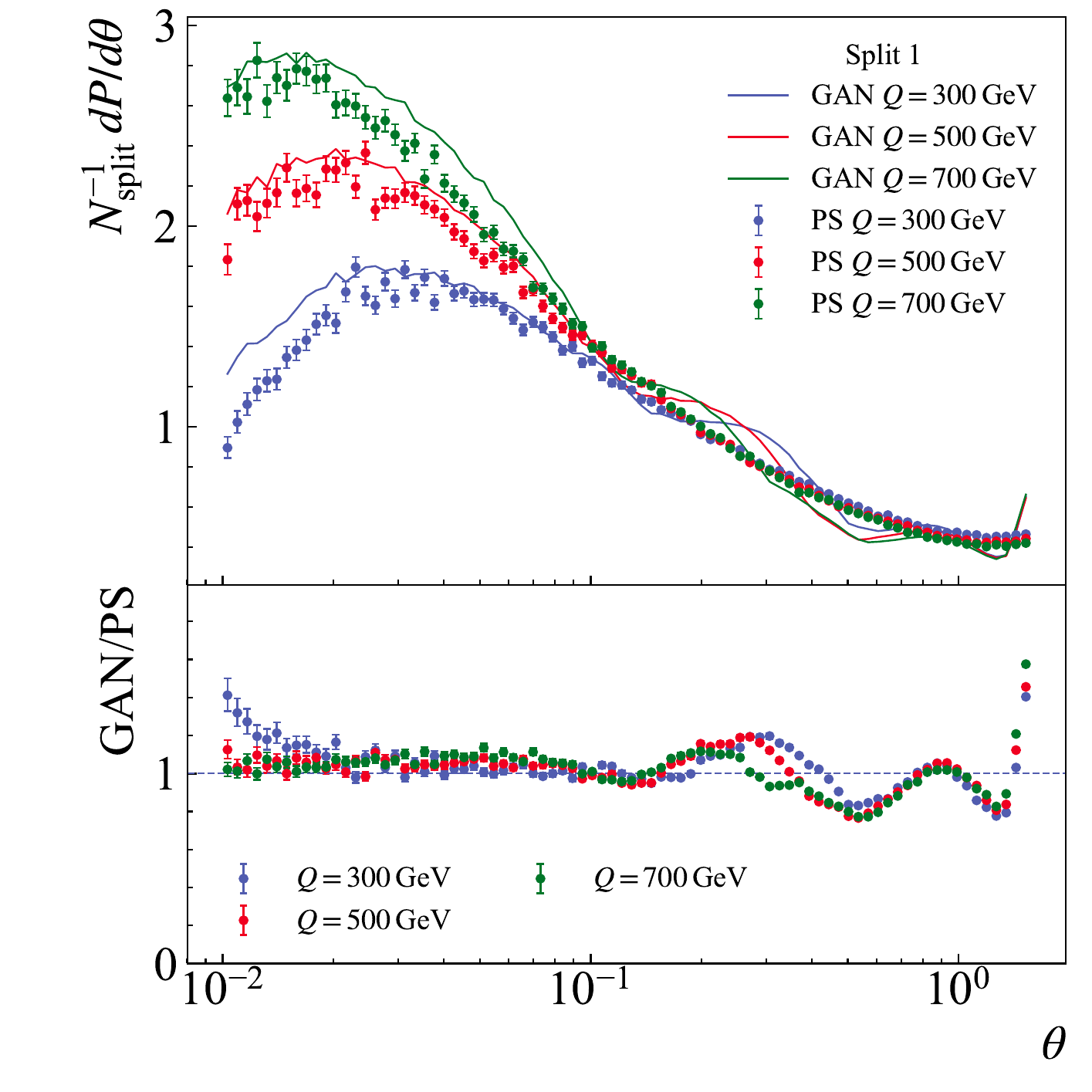}}
  \caption{Comparison of the momentum fraction $z$, i.e. the Altarelli-Parisi splitting function $P_{g\to gg}(z)$ (left) and the relative splitting angle $\theta$ (middle) of the first four splittings from the parton shower and the \gann\ for $Q=200-800$~GeV. In addition, we show the $\theta$ distribution for three different values of $Q$ for the first splitting (right).~\label{fig:splitting_dist}}
\end{figure*}

{\it Data representation and setup of the GAN.} To avoid any loss of information, we choose to train the \gann\ directly on sets which contain the event-by-event particle four-vectors produced by the shower. The required permutation invariance is built into the discriminator network by using so-called deep sets which were developed in Refs.~\cite{DBLP:journals/corr/ZaheerKRPSS17,DBLP:journals/corr/abs-1901-09006,JMLR:v21:19-322}. Several equivariant layers are followed by a permutation invariant layer which ensures that the discriminator network is insensitive to the ordering of the input. Since the number of particles that are produced per event fluctuates, the sets of four-vectors have variable length. Deep sets are ideally suited to handle input with different lengths. To accommodate the variable length of the training data we allow the deep sets to contain up to 200 four-vectors which is sufficient for the energy $Q$ that we consider here. We note that it is also possible to train the network on a set of observables where Infrared-Collinear safety is built in directly~\cite{Komiske:2018cqr,Dolan:2020qkr}. We plan to explore the impact of different data representations in future work which will be particularly relevant once we include nonperturbative effects in the shower.

The generator network mimics the structure of a parton shower. It sequentially produces partons and learns to map $n$ to $n+1$ partons. To simplify the training process, the generator is separated into a Monte Carlo time-dependent and time-independent part. The time-independent part is designed to learn the Altarelli-Parisi splitting function $P_{i\to jk}$ and the azimuthal angle $\phi$ which are the same for every branching process and independent of $Q$. Whereas the other part of the network depends on the Monte Carlo time $t$ and on the energy $Q$, i.e. it changes at every step of the shower and produces emissions which are ordered in the splitting angle $\theta$, see Eq.~(\ref{eq:MCtime}). Both parts of the generator consist of neural networks with $5$ hidden layers and $50$ neurons, which is illustrated schematically in Fig.~\ref{fig:generator}. We use the exponential linear unit (ELU)~\cite{Clevert2016FastAA} as the activation function, to avoid step functions in the resulting $z$ and $\theta$ distributions. We note that the two shower cutoffs discussed above are also explicitly included in the generator network. However, in general, we expect that the cutoffs can be chosen as trainable parameters as well.

Using the shower setup described above, we generate training data for different energies in the range of $Q=200$--$800$~GeV. As a proof of concept, we study a pure gluon shower where the gluon that splits is chosen at random. The training process of the \gann\ is a modified version of the original \gann\ approach. More details are given in the supplemental material.

{\it Numerical results.} We first verify that the \gann\ can reproduce the final distribution of particles and we then consider the underlying physics by sampling from the different layers of the network. To quantify the agreement between the shower and the \gann, we consider three kinematic variables $(Z,\Theta,\Phi)$ which characterize the final distribution of particles. The result of the \gann\ and the parton shower is shown in the three panels of Fig.~\ref{fig:final_particle_dist}, where $3.5\times 10^8$ events from the \gann\ after $700$ training epochs is compared to $3.5\times 10^7$ parton shower events. We observe very good agreement for all three distributions. The good agreement over several orders of magnitude is highly nontrivial even without considering the underlying physics. As expected for a \dglap\ shower, the distribution of the parton momentum fractions rises steeply toward small-$Z$ (left panel). The distribution of the polar angle $\Theta$ peaks in the direction of the initial parton and $\Phi$ is flat which is consistent with the flat sampling of $\phi$ for each individual splitting.

Having confirmed that the \gann\ can reproduce the final output of the parton shower, we are now going to analyze the individual splitting processes to verify that the network has also correctly learned the underlying physics. The ability of the \gann\ to extract information about parton branching mechanism is the main novelty of our work. By sampling from different layers of the network, we study the distribution of the variables $(z,\theta)$ that characterize the individual splitting processes. As representative examples, we show the results for the first four splittings in the left and middle panel of Fig.~\ref{fig:splitting_dist}. The distribution of the momentum fraction $z$ is shown in the left panel for the $g\to gg$ splitting process. We observe very good agreement with the Altarelli-Parisi splitting function $P_{g\to gg}$ for all four splittings. In particular, we note that the splitting function diverges for $z\to 1$. Instead, the final $Z$-distribution (left panel in Fig.~\ref{fig:final_particle_dist}) falls off steeply toward $Z\to 1$ as expected for a QCD fragmentation spectrum. The strikingly different behavior of the two distributions near the end point clearly demonstrates that the \gann\ has in fact learned the underlying physics mechanism. Next we consider the Monte Carlo time-dependent $\theta$ distribution which is shown in the middle panel of Fig.~\ref{fig:splitting_dist}. We observe that it is correctly reproduced by the \gann\ besides small fluctuations in the tail. The distributions peak at small values of $\theta$. As expected for the ordering variable of the shower, the distributions become more narrow for splittings that occur at later Monte Carlo time. Here, $\theta$ is the only variable that depends on the scale $Q$. We investigate its $Q$ dependence by considering the first splitting of the shower which is shown in the right panel of Fig.~\ref{fig:splitting_dist}. Even though the \gann\ is optimized to reproduce only the $Q$-integrated distribution, the $Q$-dependence of the shower is nevertheless well described by the network. We attribute the remaining numerical differences to the finite number of neurons in combination with the activation function and their ability to approximate a steep multi-differential distribution. This can be mitigated by extending the size of the neural network. Lastly, we find that the distribution of the azimuthal angle $\phi$ (not shown) also agrees with the parton shower result and we thus conclude that the \gann\ has in fact accurately learned the underlying physics of the parton shower.

{\it Conclusions.} In this letter we proposed an explainable machine learning - a White Box AI - framework which successfully learns the underlying physics of a parton shower - a hallmark of modeling high-energy particle collisions. As a proof of concept, we demonstrated that Generative Adversarial Networks (\ganns) using the full event information are capable of learning the parton cascade as described by a parton shower implementing \dglap\ evolution equations. As input to the adversarial training process we used deep sets which yield a permutation invariant representation of the training data of variable length. We found that not only the final distribution of partons in the event can be described by the network but also the physics of individual splittings processes are correctly learned by the \gann. We consider our work as a starting point of a long-term effort with the goal to eventually train networks directly on experimental data designed for extracting the underlying physics using full event information registered in the detectors. We note that the precision of our approach in falsifying theoretical modeling is limited by the systematic experimental biases which we plan to explore in subsequent publications. We expect our results to be particularly relevant for future studies of nonperturbative physics, collective effects, and the modification of the vacuum parton shower in heavy-ion collisions or electron-nucleus collisions at the future Electron-Ion Collider.

{\it Acknowledgements.} We would like to thank Barbara Jacak, James Mulligan, Stefan Prestel, Nobuo Sato and Feng Yuan for helpful discussions. YSL, MP and FR are supported by the U.S. Department of Energy under Contract No. DE-AC02-05CH11231 and the LDRD Program of Lawrence Berkeley National Laboratory. DN is supported by the U.S. Department of Energy under Contract No. DE-AC52-06NA25396 at LANL and through the LANL/LDRD Program.

\bibliographystyle{utphys}
\bibliography{main.bib}

\newpage

\widetext
\appendix

\vspace*{.5cm}
\section{Supplemental material}

We first discuss the splitting kinematics of the \dglap\ parton branching process. In particular, we focus on setting up the full event kinematics in spherical coordinates. We then present more details of the \gann\ setup.

\section{The angles of the daughter partons relative to the parent direction}

The shower is designed to conserve the momentum in the plane orthogonal to the direction of the parent parton that splits, and also conserve the energy or the light-cone momentum components parallel to the parent direction, the two being equivalent up to power-corrections in the small splitting angle limit. Formally this does not conserve the total global transverse momentum relative to the initiating parton of the cascade, on the order of $5\sim 10 \%$ of the total energy of the jet, and necessarily builds up a total non-zero invariant mass of the final state, but does preserve the angular structure of the shower, and the distribution of energy implied by the \dglap\ evolution equations. More sophisticated momentum conservation schemes exist, preserving more of the structure of the distribution of partons in phase-space. This is necessary for the resummation of logarithms beyond leading logarithmic order, but such complications are unnecessary for our proof-of-concept. 

\begin{figure}[t]
\vspace*{.7cm}
\centering
\includegraphics[width=0.4\textwidth]{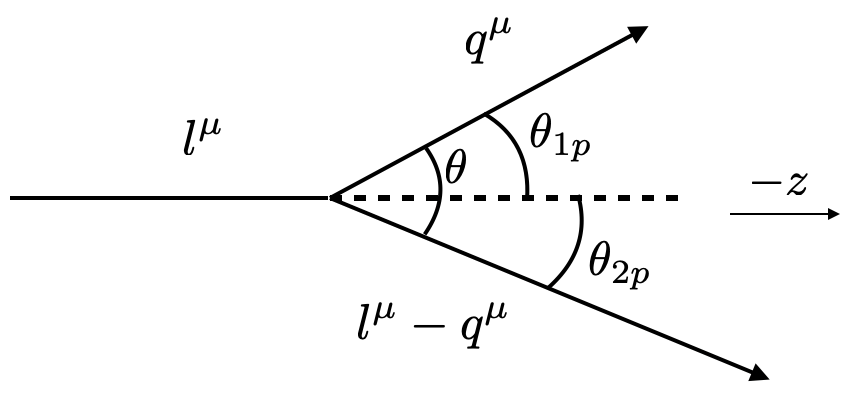}
\caption{Illustration of the \dglap\ $1\to 2$ parton branching process with the relative opening angle $\theta$ of the two daughter partons and their angles relative to the parent direction $\theta_{1,2p}$.~\label{fig:phi1+phi2}}
\end{figure}

We consider the \dglap\ $1\to 2$ parton splitting as illustrated in Fig.~\ref{fig:splitting}. The splitting process is characterized in terms of the longitudinal momentum fractions $z$ and $1-z$ of the two daughter partons, their relative opening angle $\theta$ and their orientation in azimuth $\phi$. In order to determine the spherical coordinates of the two daughter partons, we start by calculating their angle with respect to the parent direction, which we denote by $\theta_{1,2p}$. The angles $\theta_{1,2p}$ are illustrated in Fig.~\ref{fig:phi1+phi2}, and we have $\theta=\theta_{1p}+\theta_{2p}$. The two angles can be determined from the relative splitting angle $\theta$ which is related to the Monte Carlo time and the momentum fraction $z$. We consider the splitting of a parent parton with momentum $l^\mu$ (in the $-z$ direction) to two daughter partons with momentum $q^\mu$ and $l^\mu-q^\mu$. Both partons after the splitting are on-shell $q^2=(l-q)^2=0$. Using light cone coordinates, we have
\begin{equation}\label{eq:vecq}
    |\vec q\,|=q^0=\frac12(q^-+q^+)=\frac{1}{2}(zl^-+(1-z)l^+)\approx\frac12 z l^-\,.
\end{equation}
where we used
\begin{equation}
    q^+=\frac{l^+}{l^-}(l^--q^-)\,,
\end{equation}
which follows from $(l-q)^2=0$. In addition, we have $l^2=l^+l^-$ and $q^-=zl^-$. The approximation in Eq.~(\ref{eq:vecq}) holds for $l^+\ll l^-$. Similarly, we find
\begin{equation}
    |\vec l-\vec q\,|= l^0-q^0=\frac12(l^++l^--(q^++q^-))=\frac12((1-z)l^-+zl^+)\approx\frac12 (1-z)l^-\,.
\end{equation}
In order to write the angle $\theta_{1p}$ of the daughter parton with momentum $q^\mu$ in terms of the splitting angle $\theta$ and the momentum fraction $z$, we consider
\begin{equation}
    \cos\theta_{1p} = \frac{\vec q\cdot \vec l}{|\vec q\,| |\vec l\,|} \,.
\end{equation}
We rewrite the expression in terms of the momenta $q^\mu$ and $l^\mu-q^\mu$ as
\begin{align}    
    \cos\theta_{1p}=&\,\frac{\vec q\cdot (\vec l-\vec q)+|\vec q\,|^2}{|\vec q\,| |(\vec l-\vec q)+\vec q\,|} \nn\\
    =&\,\frac{|\vec q\,||\vec l-\vec q\,|\cos\theta+|\vec q\,|^2}{|\vec q\,|\sqrt{|\vec l-\vec q\,|^2+|\vec q\,|^2+2|\vec l-\vec q\,||\vec q\,|\cos\theta}}\nn\\    
    =&\, \frac{(1-z)\cos\theta_{1p}+z}{\sqrt{(1-z)^2+z^2+2z(1-z)\cos\theta}} \,.
\end{align}
The last line is obtained by inserting the expressions for $|\vec q\,|$ and $|\vec l-\vec q\,|$ which were obtained above. 
We thus find the following expression for the angle between the parent direction and the daughter parton with momentum $q^\mu$:
\begin{equation}
    \theta_{1p}=\arccos\bigg(\frac{z+(1-z)\cos\theta}{\sqrt{1-2z(1-z)(1-\cos\theta)}}\bigg)\,.
\end{equation}
Then the angle of the other daughter parton is given as $\theta_{2p}=\theta-\theta_{1p}$.

\section{The direction of the two daughter partons in absolute spherical coordinates}

\begin{figure}[t]
\vspace*{.7cm}
\centering
\includegraphics[width=0.55\textwidth]{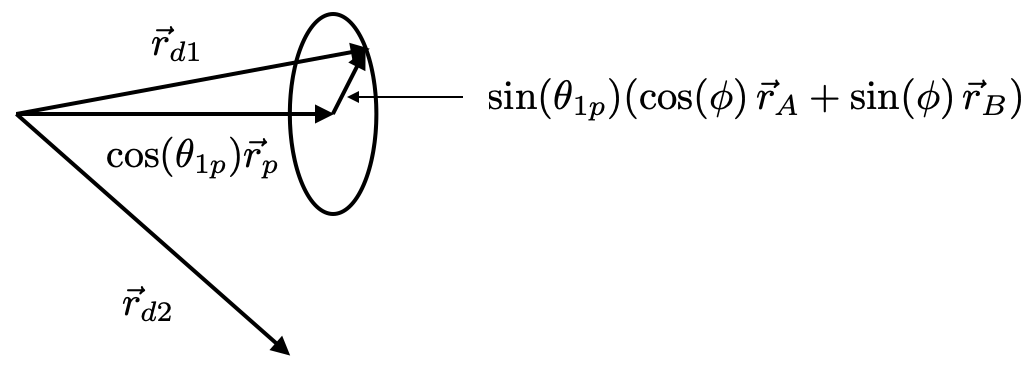}
\caption{Illustration of the vectors and angles relevant to determine the direction of the two daughter partons in absolute spherical coordinates.~\label{fig:splitting_directions}}
\end{figure}

Given the direction of the parent parton in absolute spherical coordinates $(\tilde \Theta_p,\tilde \Phi_p)$ and the kinematics of the $1\to 2$ splitting (the azimuthal direction and the angles $\theta_{ip}$ derived above), we can now determine the spherical coordinates of the two daughter partons $(\tilde \Theta_{di},\tilde \Phi_{di})$, $i=1,2$. We start with the vector pointing in the direction of the parent parton. In spherical coordinates, we have
\begin{equation}
    {\vec r}_p=
    \begin{pmatrix}
    \sin\tilde\Theta_p\cos\tilde \Phi_p \\
    \sin\tilde\Theta_p\sin\tilde\Phi_p \\
    \cos\tilde\Theta_p
    \end{pmatrix} \,.
\end{equation}
In order to generate the random distribution in azimuth, we construct a random vector $\vec r_r$ which is then orthonormalized to get a basis vector in the plane transverse to the parent direction. We use flat sampling for each component $\vec r_r^{\;i}$ in the range of $[1,-1]$. The normalized random vector transverse to the parent direction can then be written as
\begin{equation}
    \vec r_{A}=\frac{1}{N}((\vec r_p\cdot \vec r_r)\vec r_p-\vec r_r) \,,
\end{equation}
where the normalization factor $N$ is given by
\begin{equation}
    N=\left(\sum_i \left( (\vec r_p\cdot\vec r_r)^2 \vec r_p^{\;i}-\vec r_r^{\; i}\right)\right)^{1/2} \,.
\end{equation}
By construction, we thus have
\begin{equation}
    \vec r_A\cdot \vec r_p=0\,,\quad \vec r_A^{\;2}=1\,.
\end{equation}
We can then construct a second basis vector $\vec r_B$ by calculating the cross product
\begin{equation}
    \vec r_B=\vec r_A\times \vec r_{p} \,,
\end{equation}
which is normalized and orthogonal to both $\vec r_A$ and $\vec r_p$. We write the vectors $\vec r_{di}$ of the two daughter partons $i=1,2$ as a sum of two vectors. The first term is the projection of the daughter's direction onto the direction of the parent parton which is proportional $\sim\cos\theta_{1,2p}$. The second vector is in the transverse plane relative to the parent direction and parametrized in terms of $\vec r_{A,B}$ and a random variable $\phi$ chosen in the range of $[0,2\pi]$ (flat sampling). The magnitude of that second vector is given by $\sin\theta_{1,2p}$. For the two daughters $i=1,2$, the resulting vector can be written as
\begin{equation}
    \vec r_{di}=\cos(\theta_{ip}) \, \vec r_p\pm\sin(\theta_{ip})(\cos(\phi)\,\vec r_A+\sin(\phi)\,\vec r_B) \,.
\end{equation}
See Fig.~\ref{fig:splitting_directions} for an illustration of the vectors and angles relevant for setting up the full splitting kinematics of the two daughter partons. We can then write the polar and azimuthal angle of the two daughter partons as
\begin{align}
    \tilde\Theta_{di}&=\,\arccos(r_{di}^z) \,, \nn \\
    \tilde\Phi_{di}&=\,\pi+\arctan\left(\frac{r_{di}^y}{r_{di}^x}\right) \,.
\end{align}

\section{More details of the \gann\ setup}

For practical purposes we split the generator network into a time dependent and a time independent part. Both parts consist of five hidden layers with $50$ neurons and an exponential linear unit (ELU)~\cite{Clevert2016FastAA} activation to avoid discontinuous steps in the generated distributions. The time dependent network generates the next splitting angle $\theta_{i}'$ taking as input the previous angle $\theta_{i-1}'$, the initial scale $Q$, and a uniformly distributed $[0, 1)$ random number. The time independent network generates the variables $z'_i,\phi$ taking as input uniformly distributed $[0, 1)$ random numbers. We also take the momentum of the parent parton as input to the time independent network. Through the training process, the GAN learns that this information is not necessary to generate the variables $z',\phi$. To avoid vanishing gradients, the immediate output neuron of the time independent neural network generates a transformed $z'_i = -\log(1/z_i - 1)$, which is then converted to $z_i$ that is bounded by $(0, 1)$. Similarly, the $\theta'_i$ from the output neuron of the time dependent network is converted to $\theta_i$ which is bounded by $(0, \pi/2)$.

We propagate an event record of current partons, the initial scale $Q$, and the current $\theta_i$ throughout the shower process. A random parton (pure gluon shower) is selected for the splitting process by double indexing: We first sort the current list of showered partons in descending values of $Z_i$, note its indexing order, and count the number $N$ of partons when their momenta are above the cutoff $\epsilon$ and are therefore able to split. A random parton is then chosen from the first $N$ partons, and, using the order of the sorted index, it is mapped to the event record. Since the processing is \emph{de facto} executed in parallel, we calculate the splitting of a parton even if $N = 0$ which is then reversed afterwards. The highest number of branching processes occur for $Q = 800$~GeV. In this case, our implementation on Nvidia Titan RTX reaches an execution time of $95\pm 4\:\mathrm{\mu s}$/event.

The discriminator network consists of a sequence of two deep sets networks~\cite{DBLP:journals/corr/ZaheerKRPSS17,DBLP:journals/corr/abs-1901-09006,JMLR:v21:19-322}. The first deep sets network takes the list of partons from the shower as input, and produces the per-event activation. The second deep sets network uses the output of the first network, the per-event activation, as input and produces the statistical activation for the entire batch. We augment observables derived from the deep sets with the 2nd to 5th moment of the whole batch parton momenta, in order to have a fall-back in the first training epochs, until the deep sets are fully trained. The deep sets and moments are combined by a shallow network with one layer of 20 hidden neurons.

We employ a modified training process compared to the original \gann\ which we summarize here. We use the binary cross entropy as the loss function~\cite{https://doi.org/10.1111/j.2517-6161.1958.tb00292.x} which is given by
\begin{equation}
    L = -\frac{1}{2} E_x [ \log D(x) ] - \frac{1}{2} E_c [ \log (1 - D(G(c))) ] \,,
\end{equation}
where $E$ is the expectation value, $D$ the discriminator, and $G$ the generator. The conditional vector $c$ contains both the initial parton and a sufficient amount of random numbers for the full shower. The \textsc{Adam} optimizer~\cite{Kingma2015AdamAM} is used for both the discriminator and generator, where the exponential decay rate for the first and second moment are chosen as $\beta_1 = 0.5$ and $\beta_2 = 0.999$. The learning rate is $\lambda = 5\times 10^{-4}$ for the discriminator, and $\lambda = 5\times 10^{-6}$ for the generator. To guard against generator training steps that may inadvertently deteriorate the generator, the discriminator $D$ is trained in each epoch until $D(x) > 0.5$ for parton shower result $x$, and $D(G) < D(x)$ where $G$ are the partons generated by the \gann. After each generator training step, its finite step size result is tested and reverted in case it resulted in reduced $D(G)$ scores.

The time-dependent and independent networks of the generator are first pre-trained to be in the vicinity of the physical value. We observe that at the beginning of the training, the untrained discriminator allows the generator to deviate further from the pre-trained values. After $\sim 500$ epochs the discriminator is sufficiently trained to correct the generator, and closure with the parton shower occurs after $\sim 700$ epochs. This mostly concerns the $\theta$ variable whereas $z$ is more robust. We note that this is not a general limitation but allows the training to proceed by a local minimization. An alternative that we plan to explore in the future is a global optimization in combination with a random initialization.

\end{document}